\documentclass[hyper]{prop2015}
\usepackage[english]{babel}

\usepackage{amsmath,mathrsfs}
\usepackage{amsfonts}
\usepackage{amssymb}
\usepackage{stmaryrd}
\usepackage{graphicx}
\DeclareMathAlphabet{\mathpzc}{OT1}{pzc}{m}{it}

\usepackage{empheq}
\usepackage{paralist}
\usepackage{bbm}

\newcommand{\nc}{\newcommand}
\nc{\lb}{\llbracket}
\nc{\rb}{\rrbracket}
\nc{\gl}{\llbracket}
\nc{\gr}{\rrbracket}

\newcommand{\eq}[1]{\begin{equation}
                     \begin{split} #1 \end{split}
                     \end{equation}}

\allowdisplaybreaks[2]

\category{Proceedings}
\keywords{$L_\infty$-algebras, non-commutative geometry, gauge symmetries}
\subtitle{\href{http://www.maths.dur.ac.uk/lms/109/index.html}{LMS/EPSRC Durham Symposium on Higher Structures in M-Theory}}
\title{$L_\infty$-Bootstrap Approach to Non-Commutative Gauge Theories}
\author[V. G. Kupriyanov]{Vladislav G. Kupriyanov\inst{a,}\footnote{Corresponding author e-mail:~\href{mailto:vladislav.kupriyanov@gmail.com}{\textsf{vladislav.kupriyanov@gmail.com}}}}
\address[1]{CMCC Universidade Federal do ABC, Santo Andr\'e, SP, Brazil}
\address[1]{Max Planck Institut f\"ur Physik (Werner Heisenberg Institut), F\"ohringer Ring 6, 80805 M\"unchen, Germany; Report No:~MPP-2018-293}
\address[1]{Tomsk State University, Tomsk, Russia}
\begin{acknowledgements}
We would like to thank the organisers of the `Durham Symposium on Higher Structures in M-theory' for the nice and productive atmosphere. We appreciate the helpful discussions with Ralph Blumenhagen, Max Kurkov, Matthias Traube and Patrizia Vitale. The author acknowledges the hospitality of the Naples University, Federico II.
 This work was supported by the CAPES-Humboldt Fellowship No.~0079/16-2 and CNPq Grant No.~305372/2016-5.
 \end{acknowledgements}

\shortauthors{V. G. Kupriyanov}
\begin{abstract}
  A consistent description of gauge theories on coordinate dependent non-commutative (NC) space-time is a long-standing problem with a number of solutions, none of which is free from criticism. In this work, we discuss the approach proposed in \cite{Blumenhagen:2018kwq}, based on the conjecture that any consistent gauge theory can be described in terms of the $L_\infty$-structure. Starting with a well-defined commutative gauge theory, we represent it, together with the non-commutative deformation, as a part of a bigger $L_\infty$-algebra by setting some initial brackets $\ell_1$, $\ell_2$, etc. Then, solving the $L_\infty$-relations we determine the missing brackets $\ell_n$ and close the $L_\infty$-algebra defining the NC gauge theory which reproduces in the commutative limit the original one. We provide the recurrence relations for the construction of the pure gauge algebra $L^{\rm gauge}_\infty$, using which we find an explicit form of the NC $\mathfrak{su}(2)$-like and non-associative octonionic-like deformations of the Abelian gauge transformations. The construction of the $L^{\rm full}_\infty$-algebra describing the dynamics is discussed using the example of the NC Chern--Simons theory. The obtained equations of motion are non-Lagrangian, which indicates the difference between our approach and the previous ones. 
\end{abstract}
\shortabstract
\begin{document}
\maketitle

\section{Introduction}

Usually in the physics literature we understand a gauge theory as a Lagrangian field theory invariant under a certain Lie group of local transformations. If one starts with global, e.g., $U(1)$-transformations of a complex scalar field, given by, $\Phi \to \Phi^\prime=e^{if}\Phi$, with a constant $f$ and tries to make it local admitting $f$ to be a function of coordinate $x$, we observe that $\partial_a (e^{if}\Phi)\neq e^{if}\partial_a\Phi$. To fix it we introduce a gauge covariant derivative, $\mathcal{D}_a=\partial_a+i A_a$, with $A_a$ being the gauge field with transformation law $A_a^\prime=A_a-\partial_a f$. Then from the Leibniz rule it follows that the covariant derivative of a scalar field $\Phi$ transforms like the field $\Phi$ itself, $
(\mathcal{D}_a\Phi)^\prime=e^{if}\mathcal{D}_a\Phi
$.
Having this, one may proceed to construct the interacting Lagrangian between the matter field and the gauge field as well as the Yang--Mills Lagrangian describing the dynamics of the gauge field $A_a$.

The non-commutativity of space-time is a fundamental feature which can be justified by arguments from string theory and quantum gravity, among others, see  \cite{Douglas:2001ba,Szabo:2001kg} for a review. It is introduced in the theory substituting the point-wise multiplication of functions by a star product, \begin{equation}
f\cdot g\to f\star g=f\cdot g+ \frac{i\hbar}{2}\ \Theta^{ab}\partial_a f \partial_b g+\cdots\ ,
\end{equation}
 where $\Theta^{ab}$ is the anti-symmetric bi-vector field describing the non-commutativity. In the most simple and better understood case of the canonical non-commutativity, the tensor $\Theta^{ab}$ is constant. However in general the non-commutativity is a coordinate dependent field which in some cases violates the Jacobi identity. The examples are more realistic models coming from open \cite{Schomerus:1999ug,Cornalba:2001sm,Herbst:2001ai,Blumenhagen:2018kwq} and closed \cite{Blumenhagen:2010hj,Lust:2010iy} strings. For the non-constant $\Theta^{ab}$ the Leibniz rule involving the partial derivative $\partial_a$ and the star product $\star$ is violated, since
\begin{equation}
\partial_a(f\star g)=\partial_a f\star g+f\star\partial_a g+ \frac{i\hbar}{2}\ \partial_a\Theta^{cd}\partial_c f\partial_d g+\cdots\ .
\end{equation}
Consequently the above described logic is no longer applicable, the naive substitution of the point-wise products with the star products in the Lagrangian breaks gauge invariance of the theory. 

A possibility to overcome the problem with the violation of the Leibniz rule is to substitute the partial derivative $\partial_a$ with the inner derivative, defined through the star commutator, $D_a=c[x_a,\ \cdot \ ]_\star$. For the associative star product the Leibniz rule for $D_a$ follows from the Jacobi identity, see \cite{Madore:2000en,Jurco:2000ja,Jurco:2001rq} for details. The problem with this method is the commutative limit when the star commutator vanishes. Another attempt consists in invoking Hopf-algebra techniques and constructing the deformed Leibniz rule with the help of a twist element \cite{Dimitrijevic:2003pn,Vassilevich:2006tc,Dimitrijevic:2011jg,Szabo:2006wx}. Here we should mention that the very few examples of the star product originating from the twist are known, that is why the applicability of this method is very restricted. 

In this contribution to the proceedings of the Durham Symposium on Higher Structures in M-theory we discuss a recently proposed method \cite{Blumenhagen:2018kwq} which employs $L_\infty$-algebras for the construction of the consistent non-commutative and non-associative deformations of gauge theories. In Section \ref{sec:2} we give a brief introduction to $L_\infty$-algebras and discuss its relation to gauge theories. In Section \ref{sec:3} we formulate the main ideas of the $L_\infty$-bootstrap program and apply them to the construction of the pure gauge algebra $L^{\rm gauge}_\infty$ describing the gauge variations of fields, $\delta_f A$. The explicit examples of non-commutative $\mathfrak{su}(2)$-like and non-associative octonionic-like deformations of the Abelian gauge transformations are obtained in Section \ref{sec:4}. In Section \ref{sec:5} we discuss the construction of the algebra $L^{\rm full}_\infty$ which governs the dynamics of the gauge field. In particular we obtain the leading order corrections to the non-commutative Chern--Simons theory. The conclusions and open problems are given in Section \ref{sec:6}.

\section{Gauge theories from $L_\infty$-algebras}\label{sec:2}

We start this section with a formal definition of $L_\infty$-algebras. In fact, they are generalized Lie algebras where one has not only a two-bracket, the commutator, but more general multilinear $n$-brackets with $n$ inputs
\eq{
\ell_n: \qquad \quad X^{\otimes n} &\rightarrow X \\
x_1, \dots , x_n &\mapsto \ell_n(x_1, \dots , x_n) \, , 
}
defined on a graded vector space $X = \bigoplus_m X_m$, where $m\in \mathbbm{Z}$, denotes the grading of the corresponding subspace. Each element $x\in X$, has its own degree, meaning that if ${\rm deg}(x)=p$, this element belongs to the subspace $X_p$.
The concept of the degree is essential for the definition of the products $\ell_n$. First, because these brackets are graded anti-symmetric according to,
\eq{
\label{permuting}
&\ell_n (\dots, x_1,x_2, \dots) =\\ &=(-1)^{1+ {\rm deg}(x_1) {\rm deg}( x_2)} \, \ell_n (\dots, x_2,x_1, \dots )~.
}
And second, because the result $\ell_n(x_1,\ldots,x_n)\in X_p$, with
\eq{
      p={\rm deg}\big( \, \ell_n(x_1,\ldots,x_n)\, \big)=n-2+\sum_{i=1}^n  {\rm deg}(x_i)~.
}
 The set of higher brackets  $\ell_n$ define an $L_\infty$-algebra, if they satisfy the
 infinitely many relations
\eq{\label{linftyrels}
&{\cal J}_n(x_1,\ldots, x_n):=\\
&\kern.5cm:=\sum_{i + j = n + 1 } (-1)^{i(j-1)}
\sum_\sigma  (-1)^\sigma
\, \chi (\sigma;x) \times \\
 &\kern.5cm\times\ell_j \big( \;
\ell_i (x_{\sigma(1)}\; , \dots , x_{\sigma(i)} )\, , x_{\sigma(i+1)} , \dots ,
x_{\sigma(n)} \, \big) = 0 \, .
}
The permutations are restricted to the ones with
\eq{ 
\label{restrictiononpermutation}
\sigma(1) < \cdots < \sigma(i) , \qquad \sigma(i+1) < \cdots < \sigma(n)~,
}
and the sign $\chi(\sigma; x) = \pm 1$ can be read off from \eqref{permuting}. 
The first relations ${\cal J}_n$ with $n=1,2,3,\ldots$ can be schematically written as 
\eq{ 
{\cal J}_1 &= \ell_1 \ell_1~,\\[0.1cm] 
{\cal J}_2 &= \ell_1 \ell_2 - \ell_2 \ell_1 ~, \\[0.1cm] 
{\cal J}_3 &= \ell_1 \ell_3 + \ell_2 \ell_2 + \ell_3 \ell_1 ~, \\[0.1cm] 
{\cal J}_4 &= \ell_1 \ell_4 - \ell_2 \ell_3 + \ell_3 \ell_2 - \ell_4 \ell_1 \, , 
}
from which one can deduce the scheme for the higher ${\cal J}_n$. More concretely, denoting $(-1)^{x_i}=(-1)^{{\rm deg}(x_i)}$, the first $L_\infty$-relations read
\eq{\label{ininftyrel1}
& \ell_1\big( \,  \ell_1   (x) \, \big) = 0\ , \\
&\ell_1 \big( \, \ell_2(x_1, x_2)\,\big) = \\&= \ell_2\big(\,  \ell_1 (x_1) , x_2 \, \big) + (-1)^{x_1} \ell_2\big(\, x_1, \ell_1 (x_2) \, \big) \, ,
}
revealing that $\ell_1$ must be a nilpotent derivation with respect to
$\ell_2$, i.e. that in particular the Leibniz rule is satisfied.
The full relation $ {\cal J}_3 $ reads
\eq{
  \label{ininftyrel2}
       0= &\ell_1\big(\ell_3(x_1,x_2,x_3)\, \big)+\,
            +\,\ell_2\big(\ell_2(x_1,x_2),x_3\, \big)+\\ &+(-1)^{(x_2+x_3)x_1}
     \ell_2\big(\ell_2(x_2,x_3),x_1\, \big)+\\
     &+\,(-1)^{(x_1+x_2)x_3}
     \ell_2\big(\ell_2(x_3,x_1),x_2\, \big)+\\
     &+\,\ell_3\big(\ell_1(x_1),x_2,x_3\, \big)+(-1)^{x_1}\ell_3\big(x_1,\ell_1(x_2),x_3\, \big)+\\
     &+\,(-1)^{x_1+x_2}\ell_3\big(x_1,x_2,\ell_1(x_3)\, \big)\, ,
}
and means that 
the Jacobi identity for the $\ell_2$ bracket is mildly violated by
$\ell_1$ exact expressions. For the future needs we will also write here the $ {\cal J}_4 $ relation, 
\begin{equation}    \label{ininftyrel3}
  \begin{aligned}
       0&= \ell_1\big(\ell_4(x_1,x_2,x_3,x_4)\, \big)- \\
     &-\,\ell_2\big(\ell_3(x_1,x_2,x_3),x_4\, \big)+\\ &+(-1)^{x_3x_4}
    \ell_2\big(\ell_3(x_1,x_2,x_4),x_3\, \big)+\\
   &+\,(-1)^{(1+x_1)x_2}
     \ell_2\big(x_2,\ell_3(x_3,x_1),x_2\, \big)-\\
   &-\,(-1)^{x_1}\ell_2(x_1,\ell_3(x_2,x_3,x_4)\,\big)+\\
     &+\,\ell_3 \big(\ell_2(x_1,x_2),x_3,x_4\, \big)+\\
     &+\,\ell_3 \big(\ell_2(x_1,x_2),x_3,x_4\, \big)+\\ &+(-1)^{1+x_2x_3}\ell_3 \big(\ell_2(x_1,x_3),x_2,x_4\, \big)+\\
     &+\,(-1)^{1+x_2x_3}\ell_3 \big(\ell_2(x_1,x_3),x_2,x_4\, \big)+\\
     &+\,(-1)^{x_4(x_2+x_3)}\ell_3 \big(\ell_2(x_1,x_4),x_2,x_3\, \big)-\\
     &-\,\ell_3 \big(x_1,\ell_2(x_2,x_3),x_4\, \big)-\\
     &-\,\ell_4 \big(\ell_1(x_1),x_2,x_3,x_4\, \big)-\\ &-(-1)^{x_1}\ell_4 \big(x_1,\ell_1(x_2),x_3,x_4\, \big)-\\
     &-\,(-1)^{x_1+x_2}\ell_4 \big(x_1,x_2,\ell(x_3),x_4\, \big)-\\   
     &-\,(-1)^{x_1+x_2+x_3}\ell_4 \big(x_1,x_2,x_3,\ell(x_4)\, \big)~.
\end{aligned}
\end{equation}   

The framework of $L_\infty$-algebras is quite flexible
and it has been suggested  that every classical perturbative gauge
theory (derived from string theory),
including its dynamics, is organized by an underlying $L_\infty$-structure \cite{Hohm:2017pnh}. For sure, the pure gauge algebra,
called $L_\infty^{\rm gauge}$,  of such theories satisfies
the $L_\infty$-identities.  To see this, let us assume that the field theory has a standard
type gauge structure, meaning that the variations of the fields can be
organized unambiguously into a sum of terms, each of which has a definite power
in the fields. 

First we choose only two non-trivial vector spaces as
\begin{equation}\label{fA}
\begin{aligned}
 X_0  &\quad X_{-1}\\
   f     &\quad   A_a  
\end{aligned}
\end{equation}
where physically $X_0$ corresponds to the space of gauge parameters or functions $f$, and $X_{-1}$ contains the gauge fields $A_a$.
Note that in this case $\ell_1(f)\in X_{-1}$ and can be non-zero, while $\ell_1(A)\in X_{-2}$, which is empty by now, i.e., $\ell_1(A)=0$, by construction.
In this case, the only allowed non-trivial higher bracket are the ones
with one gauge parameter $\ell_{n+1}(f, A^n )\in X_{-1}$, and two gauge parameters $\ell_{n+2}(f, g, A^n )\in X_0$. The graded symmetry in this case means
\eq{
&\ell_n (\dots, f,g, \dots) =-\ell_n (\dots ,g,f, \dots )~,\\
&\ell_n (\dots, f,A, \dots) =- \ell_n (\dots,A,f, \dots )~,\\
&\ell_n (\dots, A,B, \dots) =  \ell_n (\dots,B,A, \dots )~.
}
The non-trivial $L_\infty$-relations are ${\cal J}_{n+2}(f,g,A^n)=0$ and ${\cal J}_{n+3}(f,g,h,A^n)=0$ with ${\cal J}_{n+2}(f,g,A^n)\in  X_{-1}$ and\linebreak ${\cal J}_{n+3}(f,g,h,A^n)\in X_0$.

The gauge variations are defined in terms of the brackets $\ell_{n+1}(f,A^{n})\in X_{-1}$ as follows,
\eq{\label{var}
  \delta_{f}  A &=\sum_{n\ge 0}   {1\over n!}
      (-1)^{n(n-1)\over 2}\,
 \ell_{n+1}(f, \underbrace{ A, \dots, A}_{n \; {\rm times}} )\\&=\ell_1(f)+\ell_2(f,A)-\frac12\ell_3(f,A,A)+\cdots\, .
 }
It was shown in
\cite{Hohm:2017pnh,Berends:1984rq,Fulp:2002kk}, that the $L_\infty$-relations with two gauge parameters, ${\cal J}_{n+2}(f, g, A^n )=0$, imply the
off-shell closure of the symmetry variations 
\begin{subequations}
\eq{
\label{commurel4}
                      [\delta_{f},\delta_g] A
                      =\delta_{- C(f,g, A)} A \, ,
}
where
\eq{\label{clclosure}
C(f,g, A) & =\sum_{n\ge 0} {1\over n!}
            (-1)^{{n(n-1)\over 2}}
            \, \ell_{n+2}(f,g,  \underbrace{ A, \dots, A}_{n\;  {\rm times}})~.
} 
\end{subequations}
Here we stress that the  closure relation allows for a field dependent gauge parameter.
The Jacobi identity for gauge variations
 \eq{ \label{Gaugejacobiator4}
\sum_{\rm cycl} \big[ \delta_{f}, [  \delta_{g} ,  \delta_{h} ] \big] \equiv 0 \,  ,
}
are equivalent to the $L_\infty$-relations with three gauge parameters ${\cal J}_{n+3}(f, g,h,
A^n )=0$. Thus, we see that the action of gauge symmetries on the fundamental fields is
governed by an $L_\infty^{\rm gauge}$-algebra. 

It is remarkable that
the dynamics of the theory, i.e. the equations of motion, are also expected
to fit into an extended $L_\infty^{\rm full}$-algebra.
For this purpose one extends the vector space to $X_0\oplus X_{-1}
\oplus X_{-2}$ 
\begin{equation}
\begin{aligned}\label{fAE}
X_0 &\quad X_{-1}  \quad  X_{-2} \\
                        f     &\quad   A_a    \quad  E_a
\end{aligned}
\end{equation}
where $X_{-2}$ also contains the equations of motion, 
i.e. ${\mathpzc F}\in X_{-2}$.
Now more higher brackets, namely $\ell_{n}(A^{n})\in X_{-2}$, $\ell_{n+2}(f,E,A^{n})\in X_{-2}$, and $\ell_{n+3}(f,g,E,A^{n})\in X_{-1}$, can be non-trivial and
should satisfy the following identities
\begin{eqnarray}\label{JX2}
{\cal J}_{n+1}(f,A^n)=0\qquad \mbox{and} \qquad {\cal J}_{n+2}(f,E,A^n)=0~.
\end{eqnarray}
The higher brackets $\ell_{n}(A^{n})$ are special since
they define the equation of motion, ${\mathpzc F}=0$, where
\eq{
{\mathpzc F}:=&\underset{n \ge 1}{\sum} \;{1\over n!}
(-1)^\frac{n(n-1)}{2}\, \ell_n(A^n)\\=&
   \ell_1(A)-{1\over 2} \ell_2(A^2)-{1\over 3!} \ell_3(A^3)+\cdots\; .
}
Now the $L_\infty$-structure admits that the closure condition \eqref{commurel4}
is only satisfied on-shell, i.e. there can be terms $\ell_{n+3}
(f,g,{\mathpzc F},A^n)\in X_{-1}$
on the right hand side.
The gauge variation of ${\mathpzc F}$ reads
\eq{
\label{varF}
  \delta_{f}  {\mathpzc F}=\ell_2(f,{\mathpzc F})+ \ell_3(f,{\mathpzc F},A)-{1\over 2} \ell_4(f,{\mathpzc F},A^2)+\cdots
}
reflecting that, as opposed to the gauge field $A$, it transforms covariantly.       

It was proposed that for  writing down an action for these
equations
of motion one needs an  inner bracket
\eq{
                 \langle \ \, , \ \rangle: X_{-1}\otimes X_{-2}\to
                 \mathbbm R
}
satisfying the  cyclicity property
\eq{
\label{cyclicprop}
                \langle A_0,\ell_n(A_1,\ldots,A_n) \rangle=  \langle A_1,\ell_n(A_0,\ldots,A_n )\rangle\,
}     
for all $A_i\in X_{-1}$.          
Then, the equations of motion follow from the action
\eq{
\label{actionA}
                   S&=\underset{n \ge 1}{\sum} \;{1\over (n+1)!}
(-1)^\frac{n(n-1)}{2}\, \langle A,\ell_n(A^n)\rangle\\[0.1cm]
&=  {1\over 2}\langle A,\ell_1(A)\rangle-{1\over 3!}  \langle A, \ell_2(A^2)\rangle+\cdots\; .
}

As a most simple example we discuss the $U(1)$ gauge symmetry. In this case the only non-vanishing bracket in $L^{\rm gauge}_\infty$-algebra is $\ell_1(f)=\partial_a f$. All $L_\infty$-relations are satisfied and according to (\ref{var}) we have: $\delta_f A_a=\partial_a f$. Since the gauge symmetry is Abelian the gauge closure condition reads: $  [\delta_{f},\delta_g] A  =0$. In order to extend this $L^{\rm gauge}_\infty$-algebra to a corresponding $L^{\rm full}_\infty$ one has to fix the bracket $\ell_1: X_{-1}\to X_{-2}$, depending on the choice of the theory. For the Chern--Simons theory we set: $ \ell_1(A)= \epsilon_c{}^{ab} \, \partial_a A_b$. The first $L_\infty$-relation, $\ell_1(\ell_1(f))= \epsilon_c{}^{ab} \, \partial_a \partial_bf$, is trivially satisfied. Since, all higher brackets are vanishing there is no need to check the higher $L_\infty$-relations. The corresponding equations of motion are: ${\mathpzc F}_c:=\epsilon_c{}^{ab} \, \partial_a A_b=0$. For the Maxwell theory, $ \ell_1(A)=\Box A_{a}-\partial_a(\partial\cdot A)$, implying the equations of motion, ${\mathpzc F}_a:=\partial^b(\partial_b A_a-\partial_aA_b)=0$. The $L_\infty$-description of non-Abelian gauge theories can be found in \cite{Hohm:2017pnh}.

\section{$L_\infty$-bootstrap}\label{sec:3}

In the previous section we saw how $L_\infty$-structures give rise to gauge theories. In principle, the corresponding $L_\infty$-algebra may have an infinite number of brackets $\ell_n$, which however, are not arbitrary, since they should satisfy the $L_\infty$-relations ${\cal J}_{n}=0$. The main idea of the $L_\infty$-bootstrap approach consists in representing the original undeformed gauge theory together with a deformation as a part of a bigger $L_\infty$-structure by fixing initial brackets $\ell_1$, $\ell_2$, etc. Then solving the $L_\infty$-relations ${\cal J}_{n}=0$, one determines the missing brackets $\ell_n$, which are necessary to close the algebra  $L^{\rm new}_\infty$, corresponding to the consistent deformation of the original theory. 

To illustrate the above idea in this section we consider the non-commutative deformation of Abelian gauge algebra $L^{\rm gauge}_\infty$. In this case, as it was already discussed the original undeformed theory is determined by setting the bracket $\ell_1(f)=\partial_a f$. The non-commutative deformation is introduced through the star commutator of functions which, from the consideration of anti-symmetry, should be assigned to the bracket $\ell_2(f,g)=i[f,g]_\star$. Just for simplicity let us consider the limit of slowly varying, but not necessarily small gauge fields, i.e., we discard the higher derivatives terms in the star commutator and take $\ell_2(f,g)=-\{f,g\}$ as a (quasi)-Poisson bracket.

Having non-vanishing brackets $\ell_1(f)$ and $\ell_2(f,g)$, one has to check the $L_\infty$-relation ${\cal J}_{2}(f,g)=0$, involving yet undetermined bracket $\ell_2(f,A)$. It means that now the identity ${\cal J}_{2}(f,g)=0$ becomes an equation on $\ell_2(f,A)$. Solving this equation one may proceed to the next $L_\infty$-relation, ${\cal J}_{3}(f,g,h)=0$, and define the next bracket $\ell_3(f,g,A)$, etc. The procedure should be continued until no new bracket can be determined and all $L_\infty$-relations are satisfied. Let us see how it works in practice. 

The relation ${\cal J}_{2}(f,g)=0$ reads
\eq{\label{3.3}
  \ell_1( \ell_2 (f,g))=& -\{\overbrace{\partial_a  f}^{\in X_{-1}}, g\} -\{f, \overbrace{\partial_a g}^{\in X_{-1}}\}
    - (\partial_a \Theta^{ij})\, \partial_i f \partial_j g \\
       = &\ell_2(\ell_1(f),g)+\ell_2(f,\ell_1(g))~.
}
From which one finds
\begin{equation}
\ell_2(f,A)=-\{f,A_a\} -{1\over 2} (\partial_a \Theta^{ij})\, \partial_i f A_j~.
\end{equation}
Note that the solution is not unique, one may also set, e.g.,
\begin{equation}
\ell_2^\prime(f,A)= \ell_2(f,A)+s^{ij}_a(x)\, \partial_i f A_j~, 
\end{equation}
with $s^{ij}_a(x)=s^{ji}_a(x)$. By definition, $ \ell^\prime_2(A,f):=-\ell^\prime_2(f,A)$. The symmetry of $s^{ij}_a(x)$ implies that this choice of the bracket $\ell_2^\prime(f,A)$ also satisfies the equation (\ref{3.3}). However, the symmetric part $s^{ij}_a(x)\, \partial_i f \,A_j$ can be always ``gauged away'' by $L_\infty$-quasi-isomorphism, physically equivalent to a Seiberg--Witten map \cite{Seiberg:1999vs}, see  \cite{Blumenhagen:2018shf} for more details.

Then we have to define the bracket $\ell_3(f,g,A)$ from the identity ${\cal J}_{3}(f,g,h)=0$, which reads
\eq{
&0=\ell_2(\ell_2(f,g),h)+\ell_2(\ell_2(g,h),f)+\ell_2(\ell_2(h,f),g)+\,\label{j30}\\
&+\,\ell_3(\ell_1(f),g,h)+\ell_3(f,\ell_1(g),h)+\ell_3(f,g,\ell_1(h))\ .
}
The first line is a Jacobiator, 
\eq{
&\ell_2(\ell_2(f,g),h)+\ell_2(\ell_2(g,h),f)+\ell_2(\ell_2(h,f),g)=\\
&\kern.5cm=-\Pi^{ijk} \partial_i f\partial_j g\partial_k h~.
}
For associative non-commutative deformations we may just set $\ell_3(A,f,g)=0$, while in the non-associative case one needs non-vanishing $\ell_3(A,f,g)$ to satisfy it. We set
\begin{eqnarray}
\ell_3(A,f,g)=\frac13\Pi^{ijk} A_i\partial_jf\partial_kg\ .
\end{eqnarray}

The next step is crucial for the whole construction. We have to analyze the relation ${\cal J}_{3}(f,g,A)=0$, given by
\eq{
&0=\ell_2(\ell_2(A,f),g)+\ell_2(\ell_2(f,g),A)+\ell_2(\ell_2(g,A),f)+\,\\
&+\,\ell_1(\ell_3(A,f,g))-\ell_3(A,\ell_1(f),g)-\ell_3(A,f,\ell_1(g))\ .
}
For simplicity, we replace it with ${\cal J}_{3}(g,h,\ell_1(f))=0$, written in the form
\eq{
&\ell_3(\ell_1(f),\ell_1(g),h)-\ell_3(\ell_1(f),\ell_1(h),g)=G(f,g,h)~,\label{G3}\\
&G(f,g,h):=\ell_1(\ell_3(\ell_1(f),g,h))+\,\\
&\kern1.5cm+\,\ell_2(\ell_2(\ell_1(f),g),h)+\ell_2(\ell_2(g,h),\ell_1(f))+\\
&\kern1.5cm+\,\ell_2(\ell_2(h,\ell_1(f)),g)~.
}
By construction, the above equation is antisymmetric with respect to the permutation of $g$ and $h$. The graded symmetry of the $\ell_3$ bracket,  $$\ell_3(\ell_1(f),\ell_1(g),h)= \ell_3(\ell_1(g),\ell_1(f),h),$$ implies the identity on the left-hand-side of (\ref{G3}):
\eq{
&\ell_3(\ell_1(f),\ell_1(g),h)-\ell_3(\ell_1(f),\ell_1(h),g)+\,\\
&\,\ell_3(\ell_1(h),\ell_1(f),g)-\ell_3(\ell_1(h),\ell_1(g),f)+\,\\
&\,\ell_3(\ell_1(g),\ell_1(h),f)-\ell_3(\ell_1(g),\ell_1(f),h)\equiv0\
}
which in turn requires the graded cyclicity of right-hand-side of the (\ref{G3}),
\begin{equation}
G(f,g,h)+G(h,f,g)+G(g,h,f)=0~.\label{ccG3}
\end{equation}
The latter is nothing but the consistency condition for (\ref{G3}).

It is remarkable that the consistency condition (\ref{ccG3}) follows from the previously satisfied $L_\infty$-relations, namely ${\cal J}_{2}(f,g)=0$, and ${\cal J}_{3}(f,g,h)=0$. Indeed,
taking the definition of $G(f,g,h)$, one writes
\eq{
&G(f,g,h)+G(h,f,g)+G(g,h,f)=\\
&=\ell_2(\ell_2(\ell_1(h),f),g)+\ell_2(\ell_2(f,g),\ell_1(h))+\,\\
&\kern.5cm+\,\ell_2(\ell_2(g,\ell_1(h)),f)+\ell_2(\ell_2(\ell_1(g),h),f)+\,\\
&\kern.5cm+\,\ell_2(\ell_2(h,f),\ell_1(g))+\ell_2(\ell_2(f,\ell_1(g)),h)+\,\\
&\kern.5cm+\,\ell_2(\ell_2(\ell_1(f),g),h)+\ell_2(\ell_2(g,h),\ell_1(f))+\,\\
&\kern.5cm+\,\ell_2(\ell_2(h,\ell_1(f)),g)+\ell_1(\ell_3(\ell_1(f),g,h))+\,\\
&\kern.5cm+\,\ell_1(\ell_3(f,\ell_1(g),h))+\ell_1(\ell_3(f,g,\ell_1(h)))~.
}
Using ${\cal J}_{2}(f,g)=0$, we may push $\ell_1$ out of the brackets and rewrite it as
\eq{
&\ell_1\big[\ell_2(\ell_2(f,g),h)+\ell_2(\ell_2(g,h),f)+\ell_2(\ell_2(h,f),g)\big.+\\
&\kern.5cm+\,\ell_3(\ell_1(f),g,h)+\ell_3(f,\ell_1(g),h)+\ell_3(f,g,\ell_1(h))\big]\\
&\kern1cm=\ell_1\left[{\cal J}_{3}(f,g,h)\right]=0~.
}
Which means that the consistency condition (\ref{ccG3})\linebreak holds true as a consequence of the previously satisfied $L_\infty$-relations. Taking into account (\ref{ccG3}) one may easily check that the following expression (symmetrization in $f$ and $g$ of the right-hand-side of the eq. (\ref{G3})):
\begin{equation}
\ell_3(\ell_1(f),\ell_1(g),h)=-{1\over 6}\Big(G(f,g,h)+G(g,f,h)\Big) ~,
\end{equation}
has the required graded symmetry and solves\linebreak ${\cal J}_{3}(g,h,\ell_1(f))=0$. To the best of our knowledge for the first time the solution of the algebraic equation of the type (\ref{G3}) was proposed in \cite{Kupriyanov:2008dn}.

Setting \begin{equation}
\ell_3(A,B,f)=\left.\ell_3(\ell_1(f),\ell_1(g),h)\right|_{\ell_1(f)=A;\,\ell_1(g)=B}~,
\end{equation}
 one gets,
 \begin{subequations}
\eq{
\ell_3(A,B,f)=&-{1\over 6}\Big(G_a{}^{ijk}+G_a{}^{jik}\Big) A_i
     B_j \partial_k f+\\
    &+{1\over 6} \Pi^{ijk}(\partial_a A_i B_j \partial_k f - 
    A_i \partial_a B_j \partial_k f )-\\&-{1\over 2} \Pi^{ijk}(\partial_i A_a B_j \partial_k f - 
    A_i \partial_j B_a \partial_k f )~,
}
with
\eq{
G_a{}^{ijk}= & -
  \Theta^{im} \partial_m \partial_a \Theta^{jk} -{1\over
    2} \partial_a\Theta^{jm} \partial_m\Theta^{ki}-\\&-{1\over
    2} \partial_a\Theta^{km} \partial_m\Theta^{ij}+{1\over 3}\partial_a \Pi^{ijk}~.
}
\end{subequations}
At this point we would like to stress two main things:
\begin{enumerate}[i)]
\item The consistency condition (graded cyclicity) (\ref{ccG3}) holds true as a consequence of the $L_\infty$-construction.
\item Even in the associative case one needs higher brackets to compensate the violation of the Leibniz rule.
\end{enumerate}

\subsection{Higher relations}

Once the brackets $\ell_3(f,g,A)$ and $\ell_3(f,A,B)$ are determined we may proceed to the next $L_\infty$-relation and find the brackets with four entries, $\ell_4$. First we analyze ${\cal J}_{4}(f,g,h,A)=0$, which we rewrite in the form ${\cal J}_{4}(f,g,h,\ell_1(k))=0$. Taking onto account (\ref{ininftyrel3}) we write it explicitly as:
\begin{subequations}
\eq{\label{F4}
&\ell_4(\ell_1(f),g,h,\ell_1(k))+\ell_4(f,\ell_1(g),h,\ell_1(k))\,+\\
&\kern.5cm+\,\ell_4(f,g,\ell_1(h),\ell_1(k))=F(f,g,h,k)~,
}
with
\eq{
 &F(f,g,h,k)=\\
 &\kern.5cm\ell_2(\ell_3(f,g,\ell_1(k)),h)+\ell_2(g,\ell_3(f,h,\ell_1(k)))\,-\\
  &\kern.5cm-\, \ell_2(f,\ell_3(g,h,\ell_1(k)))+\ell_3(\ell_2(f,g),h,\ell_1(k))\,-\\
  &\kern.5cm-\,\ell_3(\ell_2(f,h),g,\ell_1(k))+\ell_3(\ell_2(f,\ell_1(k)),g,h)\,-\\
  &\kern.5cm-\,\ell_3(f,\ell_2(g,h),\ell_1(k))+\ell_3(f,\ell_2(g,\ell_1(k)),h)\,+\\
 &\kern.5cm+\,  \ell_3(f,g,\ell_2(h,\ell_1(k)))~.
}
\end{subequations}
The explicit form is
\begin{equation}
F(f,g,h,k)=F^{ijkl}\,\partial_if\partial_jg\partial_kh\partial_lk~,
\end{equation}
where
\begin{equation}\label{33}
\begin{aligned}
  3F^{ijkl}&=\Theta^{km}\partial_m\Pi^{ijl}+\Theta^{jm}\partial_m\Pi^{kil}+\Theta^{im}\partial_m\Pi^{jkl}\,+\\
&\kern.3cm +\, \Pi^{kml}\partial_m\Theta^{ij}+ \Pi^{jml}\partial_m\Theta^{ki}+\Pi^{iml}\partial_m\Theta^{jk}\,+\\
&\kern.3cm+\,  \frac{1}{2}\Pi^{ijm}\partial_m\Theta^{kl}+\frac{1}{2}\Pi^{kim}\partial_m\Theta^{jl}+\frac{1}{2}\Pi^{jkm}\partial_m\Theta^{il}~.
\end{aligned}
\end{equation}
Here we follow \cite{Kupriyanov:2018yaj} for the solution of the algebraic equation (\ref{F4}). By the construction $F(f,g,h,k)$ is antisymmetric in the first three arguments and the graded symmetry of $\ell_4(\ell_1(f),g,h,\ell_1(k))$ implies the graded cyclicity (consistency condition) for $F(f,g,h,k)$, which now reads:
\eq{
&F(f,g,h,k)-F(k,f,g,h)+\\ &\kern.5cm+F(h,k,f,g)-F(g,h,k,f)=0\,.\label{ccF4}
}

Again, the consistency condition (\ref{ccF4}) holds true as a consequence of the previous $L_\infty$-relations, graded symmetry and multilinearity of the brackets $\ell_n$. As previously the solution of (\ref{F4}) is constructed by taking the corresponding symmetrization of the right-hand-side:
\begin{equation}
\ell_4(\ell_1(f),g,h,\ell_1(k))=\frac{1}{8}\left(F(f,g,h,k)+F(k,g,h,f)\right)~.
\end{equation}
Then, setting
\begin{equation}
\ell_4(A,g,h,B)=\left.\ell_4(\ell_1(f),g,h,\ell_1(k))\right|_{\ell_1(f)=A;\,\ell_1(g)=B}
\end{equation}
we conclude that
\eq{
  &\ell_4(A,g,h,B)=\\
  &\kern.5cm=\left[\frac{1}{16}\Pi^{jlm}\partial_m\Theta^{ki}+\frac{1}{16}\Pi^{jkm}\partial_m\Theta^{li}\,-\right.\\
  &\kern1cm-\,\frac{1}{16}\Pi^{ilm}\partial_m\Theta^{kj}-\frac{1}{16}\Pi^{ikm}\partial_m\Theta^{lj} \label{35a}\,-\\
  &\kern1cm-\,\left.\frac{1}{24}\Theta^{km}\partial_m\Pi^{ijl}-\frac{1}{24}\Theta^{lm}\partial_m\Pi^{ijk}\right] \partial_ig\partial_jfA_kB_l~.
}

To compete the picture at this order let us also consider the $L_\infty$-relation: ${\cal J}_{4}(f,g,A,B)=0$, which we replace with ${\cal J}_{4}(f,g,\ell_1(h),\ell_1(k))=0$, and write in the form of the equation:
\eq{\label{ell_4G}
&\ell_4(\ell_1(f),g,\ell_1(h),\ell_1(k))-\ell_4(f,\ell_1(g),\ell_1(h),\ell_1(k))=\\
&\kern.5cm=G(f,g,h,k)~,
}
where
\eq{
 &G(f,g,h,k)=\\
 &\kern.5cm=\ell_1(\ell_4(f,g,\ell_1(h),\ell_1(k))-\,\\
 &\kern1cm-\,\ell_2(\ell_3(f,g,\ell_1(h)),\ell_1(k))-\,\\
 &\kern1cm-\,\ell_2(\ell_3(f,g,\ell_1(k)),\ell_1(h))+\,\\
  &\kern1cm+\,\ell_2(g,\ell_3(f,\ell_1(h),\ell_1(k)))-\,\\
  &\kern1cm-\,\ell_2(f,\ell_3(g,\ell_1(h),\ell_1(k)))-\,\\
  &\kern1cm-\,\ell_3(\ell_2(f,\ell_1(h)),g,\ell_1(k))-\,\\
  &\kern1cm-\,\ell_3(\ell_2(f,\ell_1(k)),g,\ell_1(h))-\,\\
  &\kern1cm-\,\ell_3(f,\ell_2(g,\ell_1(h)),\ell_1(k))-\,\\
  &\kern1cm-\,\ell_3(f,\ell_2(g,\ell_1(k)),\ell_1(h))+\,\\
  &\kern1cm+\,\ell_3(\ell_2(f,g),\ell_1(h),\ell_1(k))~.
}
By construction, $G(f,g,h,k)$ is antisymmetric in the first two and symmetric in last two arguments, and as a consequence of the previous $L_\infty$-relations it satisfies the graded cyclicity relation:
\begin{equation}\label{gc4}
G(f,g,h,k)+G(g,h,f,k)+G(h,f,g,k)=0~.
\end{equation}
Taking into account (\ref{gc4}) one may check that the symmetrization in the last three arguments of the right-hand-side of the eq. (\ref{ell_4G}),
\eq{
&\ell_4(f,\ell_1(g),\ell_1(h),\ell_1(k))=\\
&=\frac{1}{12}\left(G(f,g,h,k)+G(f,h,k,g)+G(f,k,g,h)\right)\,
}
has the required graded symmetry and satisfies the equation in question.

\subsection{Recurrence relations}

For the higher relations, ${\cal J}_{n+2}(g,h,A^n)=0$, we proceed in a similar way. First we substitute them by the equations ${\cal J}_{n+2}(g,h,\ell_1(f)^n)=0$, which can be represented in the form
\eq{
&\ell_{n+2}(\ell_1(f)^{n},\ell_1(g),h)-\ell_{n+2}(\ell_1(f)^{n},\ell_1(h),g)=\\
&\kern.5cm= G(f_1,\dots,f_n,g,h)~,\label{eqn}
}
where the right hand side, $G(f_1,\dots,f_n,g,h)$, is expressed in terms of the previously defined brackets 
\begin{equation}
\ell_{m+2}(\ell_1(f)^{m},\ell_1(g),h)
\end{equation}
with $m<n$. It is symmetric in the first $n$ arguments and antisymmetric in the last two by construction. The graded symmetry of $\ell_{n+2}(\ell_1(f)^{n},\ell_1(g),h)$ implies the consistency condition (since $G(f_1,\dots,f_n,g,h)$ is symmetric in the first $n$ arguments, one needs to check the cyclicity relation with respect to the permutation of the last three slots),
\eq{
&G(f_1,\dots,f_n,g,h)+G(f_1,\dots,f_{n-1},g,h,f_n)\label{ccGn}\,+\\
&\kern.5cm+\,G(f_1,\dots,f_{n-1},h,f_n,g)=0~,
}
which follows from the previous $L_\infty$-relations and can be proved by induction. 

Following \cite{Kupriyanov:2008dn} the solution of the equation (\ref{eqn}) can be constructed taking the symmetrization of the right-hand-side in the first $n+1$ arguments, i.e.,
\eq{
&\ell_{n+2}\big(\ell_1(f)^{n},\ell_1(g),h\big)=\\
&\kern.5cm=-\frac{1}{(n+1)(n+2)}\Big(G(f_1,\dots,f_n,g,h)\,+\Big.\\
&\kern1cm\Big.+\,G(f_2,\dots,f_n,g,f_1,h)+\dots+G(f_n,\dots,f_{n-1},h)\Big)~.\label{s1}
}
And finally we obtain the expression for $\ell_{n+2}(f,A^{n+1})$, substituting in the above expression all $\ell_1(f)$ with the corresponding fields $A$.

The identities with three gauge parameters  
\eq{{\cal J}_{n+3}(f,g,h,A^n)=0~,\qquad n>1~,\nonumber} are substituted by the relations ${\cal J}_{n+3}(f,g,h,\ell_1(k)^n)=0$, written in the form:
\eq{
&\ell_{n+3}(\ell_1(f),g,h,\ell_1(k)^n)+\ell_{n+3}(f,\ell_1(g),h,\ell_1(k)^n)\,+\label{Fn}\\
&\kern.5cm+\,\ell_{n+3}(f,g,\ell_1(h),\ell_1(k)^n)=F(f,g,h,k_1,...,k_n)~.
}
The right-hand-side $F(f,g,h,k_1,...,k_n)$ is antisymmetric in the first three arguments and symmetric in last $n$ arguments, and it should also satisfy the graded cyclicity relation,
\eq{
&F(f,g,h,k_1,...,k_n)-F(k_1,f,g,h,k_2,...,k_n)\,+\label{ccFn}\\
&\kern.5cm+\,F(h,k_1,f,g,k_2,...,k_n)-F(g,h,k_1,f,k_2,...,k_n)=0~,
}
 which as before follows from the previous $L_\infty$-relations, graded symmetry and multi-linearity of the brackets $\ell_n$.
The solution of (\ref{Fn}) is constructed by taking the corresponding symmetrization of right-hand-side:
\eq{
&\ell_{n+3}\big(f,g,\ell_1(h),\ell_1(k)^n\big)=\\
&\kern.5cm=-\frac{1}{n(n+2)}\Big(F(f,g,h,k_1,...,k_n)\,+\Big.\\
&\kern3cm+F(f,g,k_1,...,k_n,h\,+\label{s2}\\
&\kern3cm\Big.+\dots+F(f,g,k_n,h,k_1,...,k_{n-1})\Big)~.
}

Our aim in this proceedings is to expose a more conceptual viewpoint on the $L_\infty$-bootstrap procedure. That is why we omit here the technical details of the proof that the graded cyclicity condition at each step follows from the previously satisfied $L_\infty$-relations, leaving it for the upcoming journal paper. However, to convince the reader of the correctness of our results, we construct explicit examples of the non-trivial non-commutative and non-associative deformations of the Abelian gauge algebra in the next section using the recurrence relations (\ref{s1}) and (\ref{s2}).

\section{Examples}\label{sec:4}

The main aim of this section is to do some explicit calculations to illustrate the proposed ideas. We will work with the most simple situation taking the non-commutativity parameter $\Theta$ to be a linear function of the coordinates. We recall that we are working in the slowly varying field limit, discarding the higher derivative terms in the star commutator and replacing it by the (quasi)-Poisson bracket. This is a ``self-consistent'' approximation of non-commutativity since the main algebraic properties of the model are preserved. If we work with the NC deformations induced by an associative star product, the star commutator satisfies the Jacobi identity, just as the corresponding Poisson bracket. In this case all higher brackets with two gauge parameters vanish, so
\begin{equation}
 [\delta_{f},\delta_g] A
                      =\delta_{-i[f,g]_\star }A \, .
\end{equation}
For non-associative deformations induced by quasi-Poisson structures the non-vanishing brackets of the type\linebreak $\ell_{n+2}(f,g,A^n)$ are required to compensate the violation of the associativity.

\subsection{NC $\mathfrak{su}(2)$-like deformation}\label{sec:4.1}

As a first example we choose the non-commutativity parameter $\Theta^{ij}(x)=2\,\varepsilon^{ijk} x^k$, which correspond to the rotationally invariant $3$D NC space \cite{Hammou:2001cc,GraciaBondia:2001ct,Rosa:2012pr,Vitale:2012dz,Galikova:2013zca,Kupriyanov:2012nb}. For two dimensional NC models with rotational symmetry one may see, e.g. \cite{Gomes:2009tk}. The corresponding Poisson bracket is
\begin{equation}\label{su2}
\{f,g\}_\varepsilon=2\,\varepsilon^{ijk} x^k\ \partial_if\ \partial_jg~,
\end{equation}
implying that $\ell_{n+2}(f,g,A^n)=0$, for $n>0$. For the first two brackets with one gauge parameter one finds, 
\eq{\label{var1}
  \ell_2(f,A) &=\{A_a,f\}_\varepsilon+\varepsilon^{abc}A_b\partial_cf~,\\
\ell_3(f,A,A)&=-\frac23\left(\partial_af A^2-\partial_bfA^bA_a\right)
 }
 with  $A^2=A_bA^b$.
Then, using the recurrence relations (\ref{s1}) we observe that the brackets $\ell_{n+3}(f,A^n)$ with the odd $n$ vanish, while for even $n$ they have the structure
\eq{\label{var2}
  \ell_{n+3}(f,A^n)=\left(\partial_af A^2-\partial_bfA^bA_a\right)\chi_n(A^2)~,
 }
 for some monomial function $\chi_n(A^2)$. The combination of (\ref{var1}) and (\ref{var2}) in (\ref{var}) results in the following Ansatz for the gauge variation:
\eq{
 \delta_{f}  A_a&=\partial_af+\{A_a,f\}_\varepsilon+\varepsilon^{abc}A_b\partial_cf\,+\\
 &\kern.5cm+\,\left(\partial_af A^2-\partial_bfA^bA_a\right)\chi(A^2)~,\label{an1}
}
where the function $\chi(A^2)$ should be determined from the closure condition,
\begin{equation}\label{an2}
 [\delta_{f},\delta_g] A
                      =\delta_{\{f,g\}_\varepsilon }A \, .
\end{equation}
Let us write
\eq{
&\delta_f\left(\delta_gA_a\right)-\delta_g\left(\delta_fA_a\right)-\delta_{\{f,g\}_\varepsilon} A_a=\label{n2}\\
&\kern.1cm=\{\delta_fA_a,g\}_\varepsilon+\varepsilon^{abc}\delta_fA_b\partial_cg\,+\\
&\kern.4cm+\,\left(2\partial_agA_b\delta_fA^b-\partial_bg\delta_fA^ba_a-\partial_bgA^b\delta_fA_a\right)\chi(A^2)\,+\\
&\kern.4cm+\,\left(\partial_ag A^2-\partial_bgA^bA_a\right)\chi^\prime(A^2)2A_c\delta_fA^c\,-\\
&\kern.4cm-\,\{\delta_gA_a,f\}_\varepsilon-\varepsilon^{abc}\delta_gA_b\partial_c\,f-\\
&\kern.4cm-\,\left(2\partial_afA_b\delta_gA^b-\partial_bf\delta_gA^ba_a-\partial_bfA^b\delta_gA_a\right)\chi(A^2)\,-\\
&\kern.4cm-\,\left(\partial_af A^2-\partial_bfA^bA_a\right)\chi^\prime(A^2)2A_c\delta_gA^c\,-\\
&\kern.4cm-\,\partial_a\{f,g\}_\varepsilon-\{A_a,\{f,g\}_\varepsilon\}_\varepsilon-\varepsilon^{abc}A_b\partial_c\{f,g\}_\varepsilon\,-\\
&\kern.4cm-\,\left(\partial_a\{f,g\}_\varepsilon A^2-\partial_b\{f,g\}_\varepsilon A^bA_a\right)\chi(A^2)~.
}
After tedious but straightforward calculations we can rewrite the right-hand-side of (\ref{n2}) as
\eq{
&\left[\partial_ag\partial_bfA^b-\partial_af\partial_bgA^b\right]\times\\
&\kern.5cm\times\left(1+3\chi(A^2)+A^2\chi^2(A^2)+2A^2\chi^\prime(A^2)\right)~.\nonumber
}
That is, requiring that
\begin{equation}\label{ode}
2t\chi^\prime(t)+1+3\chi(t)+t\chi^2(t)=0~,\qquad \chi(0)=-\frac{1}{3}~,
\end{equation}
we will obtain zero on the right-hand-side of (\ref{n2}). 
The solution of (\ref{ode}) is
\begin{equation}\label{sol}
\chi(t)=\frac1t\,\left(\sqrt{t}\cot\sqrt{t}-1\right)~.
\end{equation}

Thus, we have obtained in (\ref{an1}), (\ref{sol}) an explicit form of the non-commutative $\mathfrak{su}(2)$-like deformation of the Abelian gauge transformations in the slowly varying field approximation. Following the lines described in \cite{Kupriyanov:2015uxa} this result can be generalized for non-commutative deformations along any linear Poisson structure $\Theta^{ij}(x)$.

\subsection{Non-associative octonionic-like deformation}

Now let us repeat the calculations with the quasi-Poisson structure isomorphic to the algebra of the imaginary octonions,
\begin{equation}
\{f,g\}_\eta=2\, \eta_{IJK}\, x_K\,\partial_If\,\partial_Jg \ ,\qquad I,J,K=1,\dots,7\ ,\label{oct4}
\end{equation}
where $\eta_{IJK}$ is a completely antisymmetric tensor of rank three in seven dimensions
with non-vanishing values
\eq{
&\eta_{IJK}=+1 \qquad \mbox{for}\\& IJK = 123 , \ 435, \ 471, \ 516,
\ 572, \ 624, \ 673 \ .
}
Also it is useful to write the contraction identity \cite{Kupriyanov:2016hsm},
\eq{\label{epsilon7}
\eta_{IJK}\, \eta_{LMK}=\delta_{IL}\, \delta_{JM}-\delta_{IM}\,
\delta_{JL}+\eta_{IJLM} \ ,
}
where $\eta_{IJLM}$ is a completely antisymmetric tensor of rank four in seven dimensions
with non-vanishing values
\eq{
&\eta_{IJLM}= +1 \qquad \mbox{for} \\& IJLM = 1267, \ 1346, \ 1425, \
1537, \ 3247, \ 3256, \ 4567 \ .
}
The non-commutative deformations along this type of quasi-Poisson structures are of special interest in connection to the non-geometric backgrounds in string and M-theory \cite{Gunaydin:2016axc,Kupriyanov:2017oob}.

The expression for the non-associative octonionic-like deformation of the Abelian gauge transformations for slowly varying fields reads
\eq{
 \delta_{f}  A_I&=\partial_If+\{A_I,f\}_\eta+\eta^{IJK}A_J\partial_Kf\,+\\
 &\kern.5cm+\left(\partial_If A^2-\partial_JfA^JA_I\right)\chi(A^2)~.\label{gaugePhi}
}
The difference to a previous example is that in this case, $\ell_{n+2}(f,g,A^n)\neq0$, implying the modification of the gauge closure condition. The commutator of two gauge transformations is still a gauge transformation, however with a field dependent gauge parameter, i.e., $ [\delta_{f},\delta_g] A
                      =\delta_{-C(f,g,A) }A \, ,$ with
\eq{
C(f,g,A)=&-\{f,g\}_\eta\,-\label{Coct}\\
&\kern.5cm-2\,\eta_{IJKL}\,\partial_If\,\partial_Jg\,A_K\left(\frac{\sin 2\sqrt{A^2}}{\sqrt{A^2}}\,x_L\,+\right.\\
&\kern2cm+\left.2\,\frac{\sin^2\sqrt{A^2}}{A^2}\,\eta_{LMN}A_M\,x_N\right)~.
}
For the details of this calculation see \cite{Kupriyanov:2018yaj}.

Note that a restriction of the quasi-Poisson structure (\ref{oct4}) to the three-dimensional space with coordinates $x_i$, $i=1,2,3$, results in the Poisson structure (\ref{su2}), isomorphic to the $\mathfrak{su}(2)$ Lie algebra. Since in three dimensions  the totally antisymmetric tensor $\eta_{IJKL}$ of the rank four automatically vanishes, the $C$ bracket defined in (\ref{Coct}) becomes just a $\mathfrak{su}(2)$-like Poisson structure, and the expression (\ref{gaugePhi}) transforms into (\ref{an1}).
 
\section{$L_\infty^{\rm full}$ and field dynamics}\label{sec:5}

As it was already outlined in Section \ref{sec:2},
the equations of motion governing the dynamics of the theory also can be extracted from the $L_\infty$-structure. To this end one needs to construct an extended $L_\infty^{\rm full}$-algebra defined on the vector space $X_0\oplus X_{-1}
\oplus X_{-2}$, 
where now the subspace $X_{-2}$ also contains the equations of motion. In this section we discuss the construction of  $L_\infty^{\rm full}$-algebra on the example of NC Chern--Simons theory. Also we chose the NC $\mathfrak{su}(2)$-like deformation and write the initial brackets as
\eq{
        &   \ell_1(f) = \partial_a f~,\qquad
           \ell_1(A)=\epsilon_c{}^{ab} \, \partial_a A_b \ ,\\&\ell_2(f,g)=-\theta\ \{f,g\}_\varepsilon~,
}
where we also inserted the small parameter $\theta$ in front of the Poisson bracket for the discussion of the commutative limit in the end of this section. The brackets $\ell_{n+1}(f,A^n)$ and $\ell_{n+2}(f,g,A^n)$ defining the pure gauge algebra $L^{\rm gauge}_\infty$ were determined in Section \ref{sec:4.1}. The rest brackets $\ell_n(A^n)$, $\ell_{n+2}(f,E,A^{n})$, and $\ell_{n+3}(f,g,E,A^{n})$,\linebreak should be found from the identities (\ref{JX2}).

The first non-trivial $L_\infty$-relation is
\eq{\label{7.5}
&  {\cal J}_2(f,A):=\\
&\kern.5cm:=  \ell_1(\ell_2(f,A))-\ell_2(\ell_1(f),A)-\ell_2(f,\ell_1(A))=0 ~,
}
which we substitute as previously with
\eq{
&  {\cal J}_2(f,\ell_1(g)):= \\
&\kern.5cm:= \ell_1(\ell_2(f,\ell_1(g)))-\ell_2(\ell_1(f),\ell_1(g))=0 ~.
}
Again, the graded symmetry of the $\ell_2$ bracket,\linebreak $\ell_2(\ell_1(f),\ell_1(g))=\ell_2(\ell_1(g),\ell_1(f))$, implies a consistency condition, which is satisfied trivially due to the previous $L_\infty$-relation,
\eq{
  &\ell_1(\ell_2(f,\ell_1(g)))- \ell_1(\ell_2(g,\ell_1(f)))=\\
  &\kern.5cm=\ell_1(\ell_1(\ell_2(f,g))-{\cal J}_2(f,g))\equiv0 ~.
}
One finds
\eq{
\ell_2(\ell_1(f),\ell_1(g))=\frac12 \ell_1\left(\ell_2(f,\ell_1(g))+\ell_2(g,\ell_1(f))\right) ~.
}
We stress however that contrary to Sections \ref{sec:3} and \ref{sec:4} and because of the presence of non-empty vector space $X_{-2}$ the above expression now does not fix yet the bracket $\ell_2(A,B)$, since it may contain the structures which will vanish when one substitutes the vector fields by the gradient of function, $A\to\ell_1(f)$. Let us discuss it in more details considering the example of CS theory.

In this case one calculates
\eq{\label{7.9}
&\ell_2(\ell_1(f),\ell_1(g))=-\theta\ \epsilon_c{}^{ab}\{\partial_a f,\partial_b g\}_\varepsilon\,-\\
&\kern.5cm-\frac{\theta}{2} \epsilon_c{}^{ab}\partial_a\Theta^{ij}\big(\partial_i f\partial_j\partial_b g+\partial_i g \partial_j\partial_b f\big)~.
}
From the graded symmetry we write the most general form of
\eq{\label{lab}
\ell_2(A,B)=
&-\theta\ \epsilon_c{}^{ab}\{A_a,B_b\}_\varepsilon\,+\\
&+\alpha\ \theta\ \epsilon_c{}^{ab}\partial_a\Theta^{ij}\big(A_i\partial_jB_b +B_i  \partial_j A_b \big)\,+\\
&+\beta\ \theta\ \epsilon_c{}^{ab}\partial_a\Theta^{ij}\big(A_i\partial_bB_j +B_i  \partial_b A_j \big)
}
with $\alpha$ and $\beta$ being yet undetermined coefficients. 
From (\ref{7.9}) on finds that $-\beta=\alpha+1/2$, while the coefficient $\alpha$ now can be determined from (\ref{7.5}). To do it let us write separately  
\eq{
&\ell_1(\ell_2(f,A))=-\theta \ \epsilon_c{}^{ab}\{\partial_a f,A_b\}_\varepsilon\,-\\
&\kern.5cm-\theta \ \{f,\epsilon_c{}^{ab}\partial_a A_b\}_\varepsilon-\theta \ \epsilon_c{}^{ab} \partial_a\Theta^{ij}\partial_i f\partial_j A_b\,-\\
&\kern.5cm-\frac{\theta}{2} \epsilon_c{}^{ab} \partial_b\Theta^{ij}\partial_i \partial_af A_j- \frac{\theta}{2} \epsilon_c{}^{ab} \partial_a\Theta^{ij}\partial_i f\partial_a A_j~,
}
and from (\ref{lab}),
\eq{
&\ell_2(\ell_1(f),A)=-\theta \ \epsilon_c{}^{ab}\{\partial_a f,A_b\}_\varepsilon\,+\\
&\kern.5cm+\alpha\ \theta\  \epsilon_c{}^{ab}\partial_a\Theta^{ij}\partial_if\partial_jA_b -\frac{\theta}{2}  \epsilon_c{}^{ab}\partial_a\Theta^{ij}A_i\partial_b\partial_jf\,-\\
&\kern.5cm -\theta \left(\alpha+\frac12\right)\epsilon_c{}^{ab}\partial_a\Theta^{ij}\partial_if  \partial_b A_j ~.
}
From this we can see that taking $\alpha=-1$, and $\ell_2(f,E)=-\theta\ \{f, E\}_\varepsilon$, one solves (\ref{7.5}).
The $L_\infty$-relations for $\ell_2$ for the arguments $(AA), (fE), (AE)$ are trivially satisfied, as
they lie in trivial vector spaces $X_{-3}$ and  $X_{-4}$.

The bracket $\ell_3(E,f,g)$ contributes to the
$L_\infty$-relation ${\cal J}_3(f,g,A)=0$, which however is satisfied 
without it. Therefore, we can set $\ell_3(E,f,g)=0$.
Next  we consider ${\cal J}_3(E,f,g)=0$,
\eq{
&0=\ell_2(\ell_2(E,f),g)+\ell_2(\ell_2(g,E),f)+\ell_2(\ell_2(f,g),E)\,+\\
&\kern.5cm+\ell_3(E,\ell_1(f),g)+\ell_3(E,f,\ell_1(g))\ .
}
The first line vanishes, since it is a Jacobiator. So we set $\ell_3(E,A,f)=0$. The $L_\infty$-relations, ${\cal J}_4(E,f,g,h)=0$, and ${\cal J}_4(E,A,f,g)=0$, see (\ref{ininftyrel3}) for the explicit form, are satisfied automatically and we may set $\ell_4(E,A,B,f)=0$. The same can be shown for higher brackets of the form $\ell_{n+2}(E,A^n,f)$. Thus, we conclude that the gauge variation of the field equation (\ref{varF}) in case of the associative deformation becomes
\eq{
\label{gaugeF}
  \delta_{f}  {\mathpzc F} &=\ell_2(f,{\mathpzc F})=\theta \ \{{\mathpzc F,f}\}_\varepsilon\,.
}

The missing brackets $\ell_{n}(A^n)$ should be determined from the $L_\infty$-relations ${\cal J}_n(f,A^{n-1})=0$. The next to leading order bracket $\ell_3(A,B,C)$ was found in \cite{Blumenhagen:2018kwq}. However the explicit calculations are extremely tedious, so here we provide only the resulting expression for the NC $\mathfrak{su}(2)$-like deformation of the Abelian Chern--Simons equations of motion up to order ${\cal O}(\theta^3)$:
\eq{\label{nceom}
\mathpzc F_a:=&\varepsilon^{abc}\partial_bA_c+\\+&\theta\left(\frac12\varepsilon^{abc}\{A_b,A_c\}+2A_b\partial_aA_b- A_a\partial_bA_b-A_b\partial_bA_a\right)+\\
+&\theta^2\left(\frac{1}{3}\varepsilon^{abc}A_c\partial_b(A^2)-\frac83\varepsilon^{abc}A^2\partial_bA_c\right.-\\-&2\varepsilon^{abc}A_cA_i\partial_iA_b
-\left.2\varepsilon^{ijb}A_aA_j\partial_iA_b-\{A^2,A_a\}\right)+\\+&{\cal O}(\theta^3)=0\,.
}
Let us emphasize that following the considerations of Section \ref{sec:2}, these equations are designed in a such a way that (\ref{nceom}) transforms covariantly under the NC $\mathfrak{su}(2)$-like gauge transformation (\ref{an1}), i.e., up to order ${\cal O}(\theta^3)$ the equation (\ref{gaugeF}) holds true, which also can be checked by the explicit calculation. The correct commutative limit here is evident.

An open question here is whether we can find an explicit form of $\mathpzc F_a$. In Section \ref{sec:4.1}, solving step by step the $L_\infty$-relations we were able to conjecture the form of the gauge transformation (\ref{an1}), then the conjecture was checked and the undetermined function fixed from the closure condition (\ref{an2}). Here we have quite a similar situation. Possibly solving the next $L_\infty$-relations\linebreak ${\cal J}_4(A,B,C,f)=0$, ${\cal J}_5(A,B,C,D,f)=0$, etc. one may conjecture the form of $\mathpzc F_a$ with undetermined coefficient functions. Then the coefficients can be fixed from the equation (\ref{gaugeF}).

Nevertheless we can make an important observation regarding the properties of the obtained theory already from the first orders contributions to the equations of motion (\ref{nceom}). At the end of Section \ref{sec:2}, way how to write down an action principle for the obtained from $L_\infty$-approach equations of motion was described. To this end one needs an inner product, $ \langle \ \, , \ \rangle: X_{-1}\otimes X_{-2}\to
                 \mathbbm R$, satisfying the cyclicity property. For the field theoretical models on the NC $\mathfrak{su}(2)$-like space such a product coincides with the canonical Weyl--Moyal case, see \cite{Kupriyanov:2015uxa}:
 \eq{    
\langle A, E \rangle = \int {\rm d}^3x \, \eta^{ab} A_a\, E_b~.
}
 For the action we write
\eq{
\label{actioncsone}
    S&= {1\over 2} \langle A, \ell_1(A)
      \rangle -{1\over 3!} \langle A, \ell_2(A,A) \rangle\,+\\
      &\kern.5cm+ {1\over 4!} \langle A, \ell_3(A,A,A) \rangle+{\cal O}(\theta^3)~.
}
Variation of the second term in this action over $\delta A^a$ reproduces the same structures as to first order in $\theta$ in the equations of motion (\ref{nceom}), however with coefficients different from (\ref{nceom}) which are essential for the gauge covariance of the e.o.m. The situation with the third term in the action (\ref{actioncsone}) responsible for the second order in $\theta$ term in the equations (\ref{nceom}) is much more complicated. The contribution from terms like $\{A^2,A_a\}$ or $\varepsilon^{abc}A_c\partial_b(A^2)$ simply does not appear in the action, since $A^a\{A^2,A_a\}=A_a\varepsilon^{abc}A_c\partial_b(A^2)=0$. Consequently these terms cannot be reproduced from the action (\ref{actioncsone}). 

We conclude that the equations of motion (\ref{nceom}) are non-Lagrangian. First of all this indicates the difference between our approach and the previous ones mentioned in the introduction which were searching the NC gauge theories in the class of the Lagrangian systems. Second, we need to understand the physics behind the theories we are interested in. The mathematical tools for the investigation of properties and quantization of non-Lagrangian gauge theories were discussed e.g., in \cite{Lyakhovich:2004xd,Kazinski:2005eb}.

\section{Conclusions and outlook}\label{sec:6}

The violation of the standard Leibniz rule on the coordinate dependent non-commutative spaces makes the application of the gauge principle impossible. To solve this problem we employ the $L_\infty$-framework to bootstrap the non-commutative gauge theories. This framework is quite flexible and allows one to account not only for non-commutative but also for non-associative deformations of the gauge algebra.The violation of the Leibniz rule is compensated by the presence of the higher brackets which make the gauge transformation non-linear in the fields. The violation of the Jacobi identity for the star commutator is compensated by other higher brackets which enter the closure condition for the gauge algebra. For non-associative deformation the commutator of two gauge transformations is again a gauge transformation however with a field dependent gauge parameter. 

In this paper we derived the recurrence relations for the construction of the pure gauge algebra $L^{\rm gauge}_\infty$ and used them to find an explicit form of the non-commutative $\mathfrak{su}(2)$-like and non-associative octonionic-like deformations of the Abelian gauge transformations. The same logic in principle can be applied for the construction of the algebra $L^{\rm full}_\infty$ describing the dynamics of the system. However here we appoint the technical difficulty in the reconstruction of the bracket $\ell_n(A,B,C, \dots)$ from the given $\ell_n(\ell_1(f),B,C, \dots)$, since the first one may contain some structures which will vanish after substitution $A\to\ell_1(f)$. Finding the explicit form of equations of motion for non-commutative Chern--Simons theory described in Section \ref{sec:5} is left as an open problem.

Another open issue is the construction of the $A_\infty$-structure reproducing our $L_\infty$-algebra after the graded symmetrization. Some preliminary steps in this direction were made in the appendix of \cite{Blumenhagen:2018kwq}. This $A_\infty$-algebra can be useful for the construction of the consistent interaction with the matter fields.

\bibliographystyle{prop2015}
\bibliography{allbibtex}

\end{document}